\documentclass[mathleft]{an}
\usepackage{graphicx}
\usepackage{times}
\renewcommand{\vec}[1]{\mbox{\boldmath$#1$}}
\def\gsim{\lower.4ex\hbox{$\;\buildrel >\over{\scriptstyle\sim}\;$}}
\def\lsim{\lower.4ex\hbox{$\;\buildrel <\over{\scriptstyle\sim}\;$}}

\begin{document}

\Pagespan{1}{}
\Yearpublication{2011}%
\Yearsubmission{2010}%
\Month{XX}%
\Volume{XXX}%
\Issue{XX}%

\title{Alleviation of catastrophic quenching in solar dynamo model
    \\ with nonlocal alpha-effect}

\author{L.L. Kitchatinov\inst{1,2}\fnmsep\thanks{Corresponding author:
  \email{kit@iszf.irk.ru}\newline}
\and  S.V. Olemskoy\inst{1} }

\titlerunning{Alleviation of catastrophic quenching}

\authorrunning{L.L. Kitchatinov \& S.V. Olemskoy}

\institute{Institute for Solar-Terrestrial Physics, P.O. Box 291,
Irkutsk 664033, Russia
 \and
Pulkovo Astronomical Observatory, St. Petersburg 196140, Russia}

\received{} \accepted{} \publonline{}


\abstract{The nonlocal alpha-effect of Babcock-Leighton type is not
prone to the catastrophic quenching due to conservation of magnetic
helicity. This is shown with a dynamo model, which jointly applies
the nonlocal alpha-effect, the diamagnetic pumping, and dynamical
equation for the magnetic alpha-effect. The same model shows
catastrophic quenching when the alpha-effect is changed to its local
formulation. The nonlocal model shows the preferred excitation of
magnetic fields of dipolar symmetry, which oscillate with a period
of about ten years and have a toroidal-to-polar fields ratio of
about a thousand.
   }

\maketitle

\section{Introduction}
This paper suggest a solution for the problem of so-called \lq
catastrophic quenching' of the alpha-effect of the mean-field dynamo
theory. The standard alpha-effect of helical turbulent motions
(Parker \cite{P55}; Steenbeck, Krause \& R\"a\-d\-ler \cite{SKR66})
with its standard (algebraic) quenching by a magnetic field (Moffatt
\cite{M72}; R\"udiger \cite {R74}; Roberts \& Soward \cite{RS75};
R\"udiger \& Kitchatinov \cite{RK93}) is only a part of the total
alpha-effect known in the mean-field theory. Friesh et al.
(\cite{Fea75}) were probably the first to notice that not only
helical motions but small-scale helical magnetic fields produce the
alpha-effect. This magnetic contribution was later recognised to
lead to the catastrophic quenching of the total alpha-effect
(Vainshtein \& Cattaneo \cite{VC92}; Gruzinov \& Diamond
\cite{GD94}; Blackman \& Brandenburg \cite{BB02}).

The term \lq catastrophic' is used in the sense that the argument of
the quenching function is $\mathrm{R_m}B^2/B_\mathrm{eq}^2$, not
just $B^2/B_\mathrm{eq}^2$ as in the case of a standard algebraic
quenching ($\mathrm{R_m} = \eta_\mathrm{t}/\eta$ is the ratio of
turbulent $\eta_\mathrm{t}$ to microscopic $\eta$ magnetic
diffusivity, and $B_\mathrm{eq} = \sqrt{\mu\rho}\,u'_\mathrm{rms}$
is the energy equipartition value of a magnetic field). The magnetic
Reynolds number, $\mathrm{R_m}$, in astrophysical fluids is normally
so large that even a very small mean field $B$ suppresses the
alpha-effect in the case of catastrophic quenching.

This type of magnetic quenching is related to the conservation of
magnetic helicity. Its origin is, briefly, as follows. The
large-scale magnetic fields generated by the alpha-ef\-fe\-ct
dynamos are helical. As the magnetic helicity is conserved,
small-scale magnetic fields attain helicity equal in amount and
opposite in sign to that of large-scale fields. Helical small-scale
fields produce their own magnetic alpha-effect that counteracts the
acting alpha-effect of whatever origin so that the total
alpha-effect vanishes. Detailed discussions of catastrophic
quenching can be found in literature (cf., e.g., Brandenburg \&
Subramanian \cite{BS05}).

Catastrophic quenching presents a serious problem for the cosmic
dynamo theory. The currently leading idea in resolving the problem
for the sun is the evacuation of small-scale magnetic helicity from
the solar interior by helicity fluxes (Vishniac \& Cho \cite{VC01};
Guerrero, Chatterjee \& Brandenburg \cite{GCB10}) and then from the
solar corona by coronal mass ejections (Brandenburg \cite{B09}).
However, coronal ejections can evacuate only a minor part of
magnetic helicity (Kliem, Rust \& Seehafer \cite{KRS10}).

This paper suggests another solution related to nonlocal formulation
of the alpha-effect. If the region where the toroidal field is
concentrated is spatially separated from the region where the
poloidal field is produced by the alpha-effect, the large-scale
fields  generated will not be helical, and the problem of
catastrophic quenching does not arise. To show this, we use a
numerical model of the $\alpha\Omega$-dynamo with a nonlocal
alpha-effect of the Babcock-Leighton type (cf., e.g., Dikpati \&
Charbonneau \cite{DC99}). Another important ingredient of our model
is the diamagnetic pumping of large-scale fields. The pumping
ensures that toroidal fields are concentrated near the base of the
convection zone away from the near-top region where the alpha-effect
is active. The dynamical equation for the magnetic part of the
alpha-effect is involved. This normally leads to a strong
(catastrophic) suppression of the dynamo. We actually find
catastrophic quenching when changing to a local formulation of the
alpha-effect in our model. The nonlocal model, however, does not
show any sign of catastrophic quenching. With the largest
$\mathrm{R_m} = 10^4$ we can apply, the results of the runs with the
magnetic alpha-effect included or neglected are practically the
same. The nonlocal model reproduces the main features of the solar
cycle. There are also some disagreements with observations showing
ways for model improvement.
 \section{The model}
 \subsection{Dynamical equation for $\alpha_{_\mathrm{M}}$}
Our dynamo model is based on the mean-field induction equation
\begin{equation}
  \frac{\partial{\vec B}}{\partial t} = {\vec\nabla}\times \left(
  {\vec{\cal E}} + {\vec u}\times{\vec B} -
  \eta{\vec\nabla}\times{\vec B}\right),
  \label{1}
\end{equation}
where $\eta$ is the microscopic diffusivity, $\vec u$ is the mean
velocity, and ${\cal{\vec E}} = \langle{\vec u}'\times{\vec
B}'\rangle$ is the mean electromotive force. The electromotive force
in its local formulation,
\begin{equation}
  {\vec{\cal E}} = \langle {\vec u}'\times{\vec B}'\rangle =
  (\alpha_{_\mathrm{K}} + \alpha_{_\mathrm{M}}){\vec B} + ... ,
  \label{2}
\end{equation}
includes the magnetic alpha-effect $\alpha_{_\mathrm{M}}$ together
with the st\-a\-n\-dard kinetic $\alpha_{_\mathrm{K}}$.

The heuristic dynamical equation for the $\alpha_{_\mathrm{M}}$
\begin{equation}
  \frac{\partial \alpha_{_\mathrm{M}}}{\partial t} +
  {\vec\nabla}\cdot{\vec{\cal F}} = -
  2\frac{\eta_\mathrm{t}}{\ell^2}\left(\frac{{\vec{\cal
  E}}\cdot{\vec B}}{B_\mathrm{eq}^2} +
  \frac{\alpha_{_\mathrm{M}}}{\mathrm{R_m}}\right)
  \label{3}
\end{equation}
(Kleeorin \& Ruzmaikin \cite{KR82}; Blackman \& Brandenburg
\cite{BB02}) can be formulated without specification of a particular
form of the mean electromotive force $\vec{\cal E}$; $\ell$ is the
correlation length. For a uniform mean field, however, Eq.~(\ref{2})
applies and the steady value of the total $\alpha =
\alpha_{_\mathrm{K}} + \alpha_{_\mathrm{M}}$ can be estimated to be
 \begin{equation}
   \alpha = \frac{\alpha_{_\mathrm{K}}}{1 +
   \mathrm{R_m}\frac{B^2}{B^2_\mathrm{eq}}} .
   \label{4}
 \end{equation}
This equation describes catastrophic quenching of the alpha-effect.
We shall use the Eq.~(\ref{3}) in our model of solar dynamo to see
that the dynamo is strongly suppressed when the local formulation
(\ref{2}) for the alpha-effect is applied. Note that the magnetic
alpha-effect is local by nature in contrast to the kinetic
alpha-effect that allows a nonlocal formulation (Blackman \&
Brandenburg \cite{BB02}; Brandenburg \& K\"apyl\"a \cite{BK07}).

With a nonlocal formulation for the kinetic alpha-effect, the
toroidal field may be small in the region where the alpha-effect is
active. The $\alpha_{_\mathrm{M}}$ is also small in this case and
catastrophic quenching does not happen. We shall see that this
possibility can indeed be realised in a dynamo model with a nonlocal
(kinetic) alpha-effect.
 \subsection{Dynamo equations}
Our model accounts for the diamagnetic pumping of mean fields with
an effective velocity of
\begin{equation}
  {\vec V}_\mathrm{dia} = -\frac{1}{2}{\vec\nabla}\eta_\mathrm{t} .
  \label{5}
\end{equation}
Diamagnetic pumping was predicted analytically by Ze\-l\-dovich
(\cite{Z57}) and R\"adler (\cite{R68}). The diamagnetic effect of
inhomogeneous turbulence lacks pictorial explanation but its
existence has been confirmed by direct numerical simulations
(Brandenburg et al. \cite{Bea96}; Dorch \& Nordlund \cite{DN01};
Ziegler \& R\"udiger \cite{ZR03}) and by laboratory experiment with
turbulent liquid Sodium (Spence et al. \cite{Sea07}). The
concentration of magnetic fields at the base of convection zone by
the diamagnetic pumping can be important for a dynamo (R\"udiger \&
Brandenburg \cite{RB95}; Guerrero \& de\,Gouveia\,Dal\,Pino
\cite{GG08}; Kitchatinov \& R\"udiger \cite{KR08}).

With allowance for the diamagnetic pumping, the mean electromotive
force reads
\begin{equation}
  {\vec{\cal E}} =
  -\sqrt{\eta_{_\mathrm{T}}}\ {\vec\nabla}\times\left(
  \sqrt{\eta_{_\mathrm{T}}} {\vec B}\right) +
  \alpha_{_\mathrm{M}}{\vec B} + \vec{\cal A} ,
  \label{6}
\end{equation}
where $\eta_{_\mathrm{T}} = \eta + \eta_\mathrm{t}$ is total
magnetic diffusivity and $\vec{\cal A}$ stands for the contribution
of the kinetic alpha-effect that can be written as
\begin{equation}
  \vec{\cal A} = \int\hat{\alpha}({\vec r},{\vec r}'){\vec
  B}({\vec r}')\ \mathrm{d}^3r' .
  \label{7}
\end{equation}
With $\hat{\alpha} = \alpha_{_\mathrm{K}}\delta ({\vec r} - {\vec
r}')$, we recover the local alpha-effect. The kernel function
$\hat\alpha$ for the nonlocal formulation will be specified later.

We will consider an axisymmetric dynamo in a spherical shell. The
magnetic field in this case can be written as a superposition of its
toroidal part $B$ and a poloidal field defined with a toroidal
potential $A$:
 \begin{equation}
   {\vec B} = {\vec e}_\phi B + {\vec\nabla}\times \left({\vec
   e}_\phi\frac{A}{r\sin\theta}\right) ,
   \label{8}
 \end{equation}
where standard spherical coordinates are used and ${\vec e}_\phi$ is
the azimuthal unit vector.

Normalized variables are used. Time is measured in un\-i\-ts of
$R_\odot^2/\eta_0$; $\eta_0$ is the characteristic value of total
diffusivity. The magnetic field is normalized to the field strength
$B_0$ for which nonlinear effects become essential, and the
$\alpha$-parameter - to its characteristic value $\alpha_0$. The
poloidal field potential is measured in units of
$\alpha_0B_0R^3_\odot/\eta_0$. From now on, the same notations are
kept for the normalized variables as used before for their not
normalized counterparts, except for the fractional radius $x =
r/R_\odot$ and normalized diffusivity $\eta =
\eta_{_\mathrm{T}}/\eta_0$. The normalized equation for the toroidal
field reads
 \begin{eqnarray}
    \frac{\partial B}{\partial t} &=& \frac{\cal D}{x}
    \left(\frac{\partial f}{\partial x}\frac{\partial
    A}{\partial\theta} - \frac{\partial f}{\partial\theta}
    \frac{\partial A}{\partial x}\right)
    \nonumber \\
    &+& \frac{\eta}{x^2}\frac{\partial}{\partial\theta}\left(
    \frac{1}{\sin\theta}\frac{\partial(\sin\theta
    B)}{\partial\theta}\right)
    \nonumber \\
    &+& \frac{1}{x}\frac{\partial}{\partial
    x}\left(\sqrt{\eta}\ \frac{\partial(\sqrt{\eta}\ xB)}
    {\partial x}\right) ,
 \label{9}
 \end{eqnarray}
where
 \begin{equation}
    {\cal D} = \frac{\alpha_0 \Omega R_\odot^3}{\eta_0^2}\
 \label{10}
 \end{equation}
is the dynamo number. The $\alpha\Omega$-approximation is applied to
neglect the alpha-effect in the toroidal field equation (\ref{9}).
In this equation, $f$ is the normalized rotation frequency,
\begin{equation}
  {\vec u} = {\vec e}_\phi r\sin\theta\Omega f(x,\theta ) .
  \label{11}
\end{equation}

The equation for the poloidal field with nonlocal alpha-ef\-fe\-ct
is written as
 \begin{eqnarray}
    \frac{\partial A}{\partial t} &=&
    \frac{\eta}{x^2}\sin\theta\frac{\partial}{\partial\theta}
    \left(\frac{1}{\sin\theta}\frac{\partial
    A}{\partial\theta}\right) + \sqrt{\eta}\frac{\partial}{\partial
    x} \left(\sqrt{\eta}\frac{\partial A}{\partial x}\right)
    \nonumber \\
    &+& x \sin\theta  \cos\theta \int\limits_{x_\mathrm{i}}^x
    \hat\alpha (x,x') B(x',\theta)\ \mathrm{d} x'
    + \hat{\alpha}_{_\mathrm{M}}B ,
 \label{12}
 \end{eqnarray}
where $\hat{\alpha}_{_\mathrm{M}} = x\sin\theta
(\alpha_{_\mathrm{M}}/\alpha_0)$ is the normalised magnetic
alpha-parameter. The integration in this equation is only in the
radius with the upper limit $x$. This qualitatively reflects the
fact that the nonlocal alpha-effect in some point $x$ is contributed
by the buoyant magnetic loops rising from deeper layers ($x' < x$)
of the convection zone and that buoyant velocities are almost
vertical.

The equation system is closed with the dynamical equation for
$\hat{\alpha}_{_\mathrm{M}}$. We neglect the helicity flux in
Eq.~(\ref{3}) and use the $\alpha\Omega$-approximation to rewrite
this equation in normalized units:
\begin{eqnarray}
    \frac{\partial\hat{\alpha}_{_\mathrm{M}}}{\partial t} &=&
    -2\hat{\eta}\left(\frac{R_\odot}{\ell}\right)^2\left(
    B\frac{\partial A}{\partial t}
    -\frac{\hat{\eta}}{x^2\sin\theta}\frac{\partial
    A}{\partial\theta}\frac{\partial (\sin\theta
    B)}{\partial\theta} \right.
    \nonumber \\
    &-& \left.\frac{\sqrt{\hat{\eta}}}{x}
    \frac{\partial (\sqrt{\hat{\eta}} x B )}{\partial x}
    \frac{\partial A}{\partial x}\right)
    -
    2\left(\frac{R_\odot}{\ell}\right)^2
    \frac{\hat{\alpha}_{_\mathrm{M}}}{\mathrm{R_m}} ,
    \label{13}
\end{eqnarray}
where $\hat\eta = \eta_\mathrm{t}/\eta_0 = \eta - \mathrm{R_m^{-1}}$
is the normalized turbulent diffusivity. The magnetic alpha-effect
and related catastrophic quenching can be switched off by omitting
the last term in Eq.~(\ref{12}). Later on, we will compare the
results obtained with account for the magnetic alpha effect and
without it.

The boundary conditions assume an interface with a superconductor on
the inner boundary of radius $x_\mathrm{i}$,
\begin{equation}
    \frac{\partial \left( \sqrt{\eta} x B\right)}{\partial x} = 0,\ \
    A = 0\ \ \ \mathrm{for}\ \ \  x = x_\mathrm{i} ,
    \label{14}
\end{equation}
and pseudo-vacuum conditions on the top,
\begin{equation}
    B = 0,\ \ \frac{\partial A}{\partial x} = 0\ \ \ \mathrm{for}\ \ \  x = 1.
    \label{15}
\end{equation}
 \subsection{Model design}
\begin{figure}
    \includegraphics[width=70mm]{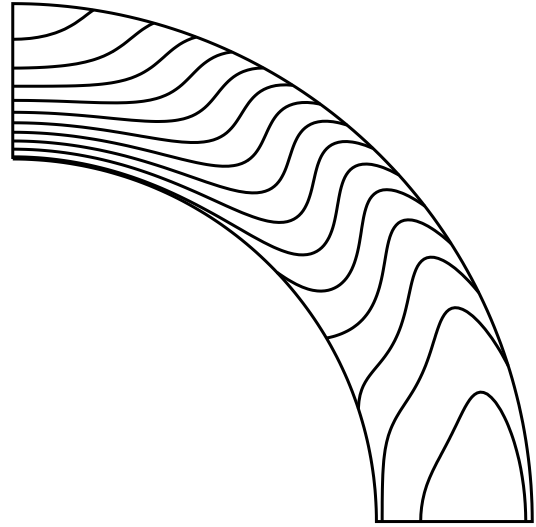}
    \caption{Angular velocity isolines for the differential rotation
    used in the dynamo model.
              }
    \label{f1}
\end{figure}

For the differential rotation, we use the approximation by
Belvedere, Kuzanyan \& Sokoloff (\cite{Bea00}) for
helioseismological data
 \begin{equation}
    f(x,\theta) = \frac{1}{461}\sum\limits_{m=0}^{2}
    \cos\left( 2m\left(\frac{\pi}{2}
    - \theta\right)\right) \sum\limits_{n=0}^{4} C_{nm}x^n .
    \label{16}
 \end{equation}
The coefficients $C_{mn}$ of this equation are given in Table~1 of
Belvedere et al. (\cite{Bea00}). Figure~\ref{f1} shows the angular
velocity isolines.

The kernel function of the nonlocal alpha-effect in the poloidal
field equation (\ref{12}) was prescribed as follows
\begin{eqnarray}
    \hat\alpha (x,x') &=& \frac{\phi_\mathrm{b}(x')\phi_\alpha (x)} {1 + B^2(x',\theta)}
    ,
    \nonumber \\
    \phi_\mathrm{b}(x') &=& \frac{1}{2}\left( 1 -
    \mathrm{erf}\left( (x' - x_\mathrm{b})/h_\mathrm{b}\right)\right) ,
    \nonumber \\
    \phi_\alpha (x) &=& \frac{1}{2}\left( 1 +
    \mathrm{erf}\left( (x - x_\alpha)/h_\alpha\right)\right),
    \label{17}
\end{eqnarray}
where $\mathrm{erf}$ is the error function and $B^2$ in the
denominator of the first equation accounts for the usual algebraic
quenching of the alpha-effect. We always use $x_\mathrm{b} =
x_\mathrm{i} + 2.5h_\mathrm{b}$ and $x_\alpha = 1 - 2.5h_\alpha$ to
ensure smoothness of the kernel functions in the simulation domain.
The $h_\mathrm{b}$-parameter is the thickness of the near-bottom
region of toroidal magnetic fields producing the alpha-effect. The
$h_\alpha$ is the thickness of the near-surface region where this
alpha-effect is produced. The nonlocal alpha-effect with the kernel
function (\ref{17}) is very similar to the Babcock-Leighton
mechanism for the poloidal field production used in the dynamo
models of Durney (\cite{D95}) and Dikpati \& Charbonneau
(\cite{DC99}).

To compare the results of the simulations for nonlocal and local
alpha-effects we will need a local formulation of this effect, which
can be obtained by applying the kernel function
\begin{equation}
    \hat\alpha (x,x') = \frac{2\hat{\eta}\delta (x - x')}{1 + B^2} .
    \label{18}
\end{equation}

\begin{figure}
    \includegraphics[width=70mm]{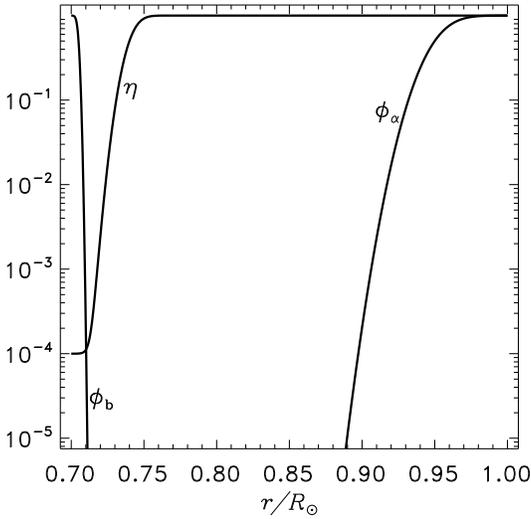}
    \caption{Profiles of the normalised total diffusivity and the
    kernel functions (\ref{17}) of the nonlocal alpha-effect.
              }
    \label{f2}
\end{figure}

The diffusivity profile of our model reads
\begin{equation}
    \eta(x) = \mathrm{R_m^{-1}} + \frac{1}{2}
    (1 - \mathrm{R_m^{-1}})\left( 1 + \mathrm{erf}\left(\frac{x -
    x_\eta}{h_\eta}\right)\right) .
    \label{19}
\end{equation}
The aim of this paper is to demonstrate that the nonlocal
alpha-effect is not prone to catastrophic quenching. We will do this
with the following parameter values: $x_\mathrm{i} = 0.7$, $x_\eta =
0.74$, $h_\eta = 0.01$, $h_\alpha = 0.02$, $h_\mathrm{b} = 0.002$,
and $R_\odot /\ell = 10$. The dependence on the model parameters
will be discussed in a separate publication. The profiles of
diffusivity and kernel functions (\ref{17}) for this parameters set
are shown in Fig.~\ref{f2}. The maximum value of the magnetic
Reynolds number we were able to apply is $\mathrm{R_m} = 10^4$, and
all the results of this paper were obtained with this value.

The system of dynamo equations (\ref{9}), (\ref{12}) and (\ref{13})
was solved numerically with the grid-point method and explicit
ti\-me-stepping. The diamagnetic pumping and low diffusion in the
near-bottom region leads to a high concentration of the magnetic
field near the bottom. To resolve the fine near-bo\-t\-t\-om
structure, a nonuniform grid over the radius with the grid spacing
$\Delta x \sim \eta^{1/2}$ was applied. The grid over the latitude
was uniform. All the results of the next Section do not depend on
numerical resolution. This was checked by repeating the runs with
doubled resolution.

Equatorial symmetry was usually not prescribed. The field was
evolved in time starting from a mixed-parity initial field and the
solution relaxed eventually to a certain equatorial symmetry. In
order to determine the critical dynamo numbers for excitation of the
dipolar ($B(\theta ) = -B(\pi-\theta)$) and quadrupolar ($B(\theta )
= B(\pi-\theta)$) dynamo modes, additional boundary conditions
selecting the field mode of certain equatorial symmetry were imposed
on the equator.
 \section{Results and discussion}
Our computations show that catastrophic quenching is present for
local formulations of the alpha-effect but it does not exist for the
model with a nonlocal alpha effect.

\begin{figure}
    \includegraphics[width=83mm]{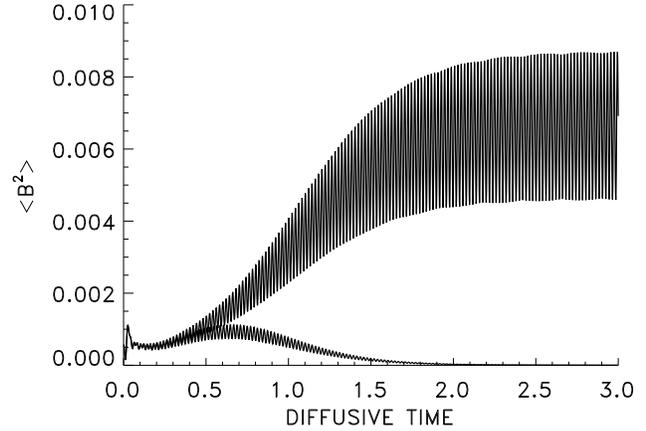}
    \caption{Time dependencies of the volume-averaged square of the toroidal field
    $\langle B^2\rangle$ for the runs with algebraic
    ($\alpha_{_\mathrm{M}} = 0$, upper line) and catastrophic ($\alpha_{_\mathrm{M}} \neq
    0$) quenching of the local alpha effect of Eq.~(\ref{18}). Time
    is measured in units of $R^2_\odot/\eta_0$. The computations
    were performed for dynamo number $D = 5.3\times 10^4$ slightly
    above the critical value of $D_\mathrm{c} = 5.0\times 10^4$.
              }
    \label{f3}
\end{figure}

Figure~\ref{f3} shows the results of computations for the local
alpha-effect of Eq.~(\ref{18}). The upper line in this Figure was
obtained with the $\alpha_{_\mathrm{M}}$ put to zero in the poloidal
field equation (\ref{12}). Saturation of the dynamo in this case is
due to the usual algebraic quenching of the alpha-effect. The lower
line represents the computation with allowance for the magnetic
alpha, $\alpha_{_\mathrm{M}}$, governed by the dynamical equation
(\ref{13}). Magnetic energy considerably decreases in this case
indicating catastrophic quenching of the dynamo.

\begin{figure}
    \includegraphics[width=83mm]{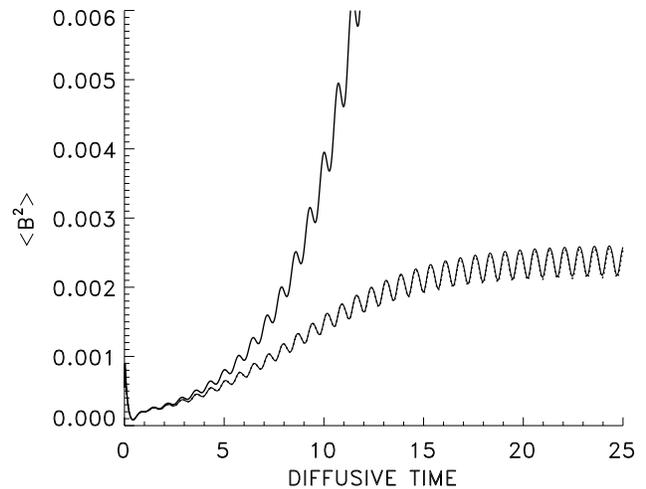}
    \caption{Time dependencies of the mean square of the toroidal field in
    the model with nonlocal alpha-effect. The two bottom lines show the
    results of the runs with $\alpha_{_\mathrm{M}} = 0$ (full line) and
    for the complete model ($\alpha_{_\mathrm{M}} \neq 0$, dotted line).
    These two lines are hard to distinguish by eye. The upper line
    is for the computation neglecting the algebraic alpha-quenching.
    The field growth does not saturate in this case in
    spite of the finite $\alpha_{_\mathrm{M}}$. The dynamo number $D =
    2.2\times 10^4$ is slightly above the critical value of
    $D_\mathrm{c} = 1.9\times 10^4$.
              }
    \label{f4}
\end{figure}

In the model with the nonlocal alpha-effect of Eq.~(\ref{17}),
saturation of the dynamo is due to the standard algebraic quenching
only. The results of Fig.~\ref{f4} for $\alpha_{_\mathrm{M}}$
neglected or included are practically the same. When algebraic
quenching is switched off by omitting $B^2$ in the denominator of
the first equation of (\ref{17}), but $\alpha_{_\mathrm{M}}$ is kept
finite, field growth does not saturate. Clearly, the
$\alpha_{_\mathrm{M}}$ does not play any role in these simulations.

\begin{figure}
    \includegraphics[width=83mm]{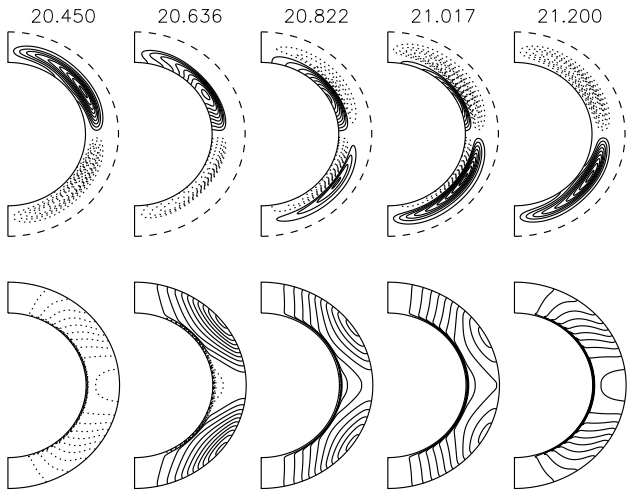}
    \caption{Toroidal field isolines (top row) and poloidal filed lines
    (bottom row) for several instances of a magnetic cycle. The time of the
    run in units of $R^2_\odot/\eta_0$ is shown at the top. Full
    (dotted) lines show positive (negative) levels and clockwise
    (anticlockwise) circulation. The pictures of the upper row are
    rescaled so that the upper (dashed) boundary shows the radius of
    $r = 0.74R_\odot$ below which the toroidal fields are localized.
    $D = 2.2\times 10^4$.
              }
    \label{f5}
\end{figure}

The reason for the inefficiency of $\alpha_{_\mathrm{M}}$ in the
nonlocal dynamo model can be seen from Fig.~\ref{f5}, which shows
the magnetic field patterns for several instances of a magnetic
cycle. The toroidal field is highly concentrated at the bottom. It
is small in the near-top region where the alpha-effect is active. In
this case, the alpha-effect does not contribute to the first term on
the right of the equation (\ref{3}) and the generated large-scale
fields are not helical. The small-scale magnetic helicity balancing
the helicity of large-scale fields is not produced and the
$\alpha_{_\mathrm{M}}$ is small.

It should be noted that the non-locality of the alpha-effect alone
does not guaranty nonoccurrence of catastrophic quenching. If the
toroidal field in the near-top region of the alpha-effect has same
sign and same order of magnitude as the near-bottom toroidal field
that produces this alpha-effect, then catastrophic quenching can
still happen (Brandenburg \& K\"apyl\"a \cite{BK07}). It can be
imagined, however, that if such a distributed toroidal field has
opposite signs near the top and near the bottom, then catastrophic
\emph{amplification} of the magnetic field will happen.

\begin{figure}
    \includegraphics[width=83mm]{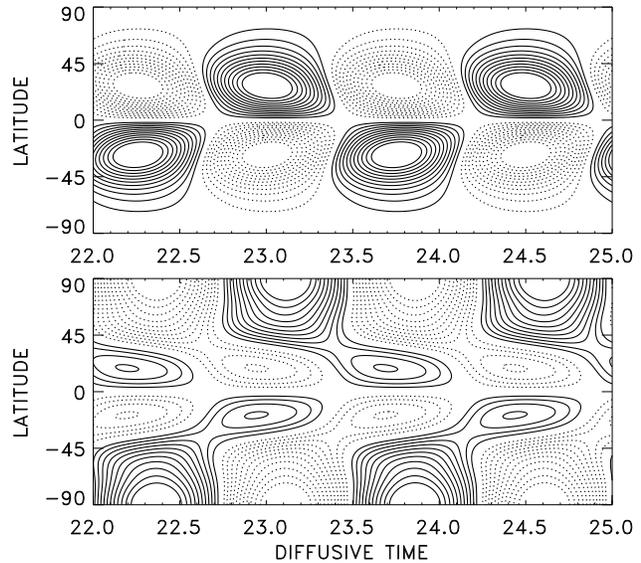}
    \caption{Butterfly diagram of the depth-integrated toroidal field
    $\cal B$ of Eq.~(\ref{20}) (top panel) and surface radial field
    (bottom) for the model with nonlocal alpha-effect. Time is shown
    in units of $R^2_\odot /\eta_0$. $D = 2.2\times 10^4$.
              }
    \label{f6}
\end{figure}

What excludes the catastrophic quenching in our model is the joint
effect of non-locality and diamagnetic confinement of the toroidal
field in the near-bottom region. This region has been long
recognised as a favorable site for the solar dynamo (cf., e.g.,
Gilman \cite{G92}). The reason why the magnetic field should be
concentrated at this site was not clear, however. We suggest that
the concentration can be accounted for by diamagnetic pumping.

The poloidal field in Fig.~\ref{f5} is also concentrated at the
bottom. The following consideration shows that somewhere inside the
sun the poloidal field must be much stronger than on the surface.
The magnitude of the surface poloidal field is about 1-2 Gauss only.
If the toroidal field is estimated by its magnitude in sunspots, it
is about 1000 times stronger than the surface poloidal field. The
solar differential rotation of about 30\% can produce in the 11
years of the solar cycle a toroidal field that is at most 40 times
stronger than the poloidal filed (the strong radial shear in the
tachocline does not change this estimation because the radial field
should be as much weaker there compared to the meridional field as
the radial shear is larger than the latitudinal shear). Therefore, a
poloidal field of several tens Gauss is necessary for the production
of kilogauss toroidal fields. The toroidal field in our dynamo model
is about 1000 times stronger than the polar field due to the
near-bottom concentration of the poloidal field.

Figure~\ref{f6} shows butterfly diagrams for the surface radial and
deep toroidal fields. The toroidal field diagram shows the isolines
of the quantity
\begin{equation}
    {\cal B} = \sin\theta \int\limits_{x_\mathrm{i}}^1
    \phi_\mathrm{b}(x) B(x)\ \mathrm{d}x ,
    \label{20}
\end{equation}
to which the Babcock-Leighton surface alpha-effect is proportional.
The factor $\sin\theta$ in Eq.~(\ref{20}) accounts for the
dependence of the length of toroidal flux tubes on latitude (it is
supposed that the probability of sunspot production is proportional
to the length of the tube).

The period of the magnetic cycles of Figures \ref{f4} and \ref{f6}
equals $0.75R^2_\odot/\eta_0$, that for the diffusivity of $\eta_0
\approx 10^{13}$\,cm$^2$/s is close to the 11 year period of the
solar cycle. The model with nonlocal alpha-effect does not suffer
from the old problem of too short magnetic cycles typical of the
local dynamo models (Fig.~\ref{3}).

The radial field diagram of Fig.~\ref{f6} is similar to
observational diagrams of Stenflo (\cite{S88}) and Obridko et al.
(\cite{Oea06}).

The field pattern of Fig.~\ref{f6} is antisymmetric about the
equator. The critical dynamo number for excitation of the dipolar
modes, $D_\mathrm{c}^\mathrm{d} = 1.9\times 10^4$, is substantially
smaller than the critical number $D_\mathrm{c}^\mathrm{q} =
2.5\times 10^4$ for the quadrupolar modes. Accordingly, the runs
started from an initial field of mixed parity rapidly relaxed to
dipolar parity. So clear preference for the dipolar modes is typical
of the dynamo models with relative small magnetic diffusion near the
base of the convection zone (Chatterjee, Nandy \& Choudhuri
\cite{CNC04}).

The only clear disagreement of the model with observations is the
presence of toroidal fields in Fig.~\ref{f6} on too high latitudes.
It may be expected that even a slow meridional flow in the
near-bottom region of low diffusion can influence the latitudinal
profile of a toroidal field. Allowance for the meridional flow is a
perspective for model improvement.
 \acknowledgements
This work was supported by the Russian Fo\-un\-dation for Basic
Research (projects 10-02-00148, 10-02-00391).

\end{document}